\documentclass[aps,prl,twocolumn,showpacs,superscriptaddress,groupedaddress]{revtex4}  % for review and submission
\usepackage{graphicx}  % needed for figures
\usepackage{dcolumn}   % needed for some tables
\usepackage{bm}        % for math
\usepackage{amssymb}   % for math
\usepackage{graphicx}
\usepackage{color}
\begin{document}

% The following information is for internal review, please remove them for submission
%\widetext
%\leftline{Version 04 as of \today}

% the following line is for submission, including submission to the arXiv!!
%\hspace{5.2in} \mbox{Fermilab-Pub-04/xxx-E}

\title{Ultrahigh Charge Carrier Mobility in Nanotube Encapsulated  Coronene Stack}
\author{Saientan Bag}
%\email{xxx}
\affiliation{Center for Condensed Matter Theory, Department of Physics, Indian Institute of Science, Bangalore-560012, India}
\author{Prabal K. Maiti}
\email{maiti@physics.iisc.ernet.in}
\affiliation{Center for Condensed Matter Theory, Department of Physics, Indian Institute of Science, Bangalore-560012, India}      
\date{\today}

\begin{abstract}
Achieving high charge carrier mobility is the holy grail of organic electronics. In this letter we report a record charge carrier
mobility of 14.93 cm$^2$ V$^{-1}$s$^{-1}$ through a coronene stack encapsulated 
in a single walled carbon nanotube (CNT) by using a multiscale modeling technique which combines MD simulations, first principle calculations and 
Kinetic Monte Carlo simulations. For the CNT having a diameter of 1.56 nm we find a highly ordered defect free  organization of coronene 
molecules inside the CNT which is responsible for the high charge carrier mobility. The encapsulated coronene molecules  are correlated with a large 
correlation length of $\sim $18 {\AA} which is independent of the length of the coronene column. Our simulation further suggests that 
coronene molecules can spontaneously enter the CNT, suggesting that the encapsulation is experimentally realizable.   
\end{abstract}

\maketitle

%section{\label{sec:level1}First-level heading}
% sections are not used for PRL papers
\begin{center} {\bf INTRODUCTION } \end{center}

The successful use of  organic electronics in everyday applications rely on the availability of materials with high
charge carrier mobility. For this purpose, there is world wide effort to design new materials which allow fast movement of charge carrier through it. 
The highest charge carrier mobility values so far reported  in the literature lie in the range of ~10-20 cm$^2$V$^{-1}$s$^{-1}$\cite{mine, diao, hsin, kang}. However the
generation of morphology to achieve such high mobility  require sophisticated experimental procedure which are still very expensive and inefficient for
large scale production.

Use of flat disk-like molecules (discotic) which are often self assembled in a columnar stack  to form   various discotic liquid 
crystalline phase is an attractive route to device  charge transporting material in organic electronicsl\cite{adam}. Utilizing the high $\pi - \pi $ overlap\cite{bag,kirk,feng} of the electronic cloud of the discotics in a 
columnar arrangement, one dimensional transport of charge through the stack is plausible. However, the efficiency of  charge transport
through the  columnar arrangement of the discotics depends on the positional and orientational correlation of the discotics and is 
limited 
by the structural defects present in a column. To maximize charge carrier mobility in a discotic assembly two
difficulties have to be overcome: First, one has to make the  columnar arrangements defect free and second, to achieve optimum positional and orientational correlations between the molecules in the column. In this race to achieve high mobility in discotics,
researchers have made significant progress in achieving a required correlation between the discotics in the column by tuning the core and
tails of the discotics.  Andrienko et. al\cite{feng} proposed a discotic material with triangular core and alternating hydrophilic/
hydrophobic side chains which self assembles in a discotic columnar phase with an average twist of $~60^{\circ}$ between the molecules. 
With this particular discotic phase, they were able to attain a mobility of only  0.2 cm$^2$V$^{-1}$s$^{-1}$ due to structural 
defects.  So the main challenge
at this stage is to find  a way to make a defect free organization of the molecules. Although the emergence of defects in the  self
assembled discotics are inevitable due to the thermal fluctuation, appearance of defects can significantly be
reduced  by suitable confinement\cite{zhang,cer,bus}. 
 
 In this letter we propose a way to achieve defect free columnar aggregates and demonstrate record charge carrier mobility through the 
column. In particular we show defect free columnar aggregate of coronene molecules inside carbon nanotube (CNT).  
While this system had been
previously studied by various groups\cite{ver,ano} in some different context, the charge transport properties of the confined coronene 
column
has not been studied so far. We use  molecular dynamics (MD) simulation to study the column formation inside CNTs of various diameter
and report a critical value of the diameter
of  the CNT to achieve a highly ordered defect free coronene stack. We use Marcus-Hush\cite{mar} formalism to
calculate the charge carrier mobility along the columnar stack. We find a record mobility of 14.93 cm$^2$V$^{-1}$s$^{-1}$  at
an electric field strength of 10$^{7}$ V/m and a temperature of 300 K. The calculated mobility value is  two order of magnitude
higher than the mobility for discotic\cite{feng} systems reported so far in the literature. 
As we have already  mentioned the highest charge carrier mobility values reported in some polymer semiconductor based organic field-effect transistors 
(OFETs)\cite{mine, diao, hsin, kang}, lie 
in the range of $\sim$ 10$-$20 cm$^2$V$^{-1}$s$^{-1}$. The mobility value that we report in this letter is as high as these.  

We also perform a simulation where the
coronene molecules were initially  placed outside the nanotube. It showed the spontaneous insertion  of the coronene molecules
inside the CNT and the formation of a well ordered column. This simulation suggests an experiment protocol where the CNT is
placed inside the coronene vapor to achieve the CNT encapsulated coronene stack. 
          
The letter is divided in three sections. In the first section we describe the MD simulation of the coronene stack. In the second section
we describe the charge transport simulation and the results. Finally we end with the concluding remarks. 
   
\begin{center} {\bf MD Simulation } \end{center}

The zigzag carbon nanotube with various radius was initially prepared using VMD\cite{hum}. The coronene molecule was built and geometry
optimized in Gaussian 09\cite{gau} using B3LYP/6-311g level of theory. The charge in the coronene molecules was calculated following
RESP method\cite{bay} using Gaussian 09\cite{gau} and Ambertools\cite{pearl} package. No charge was assigned on the carbon atom of the
nanotube. Initially  14 coronene molecules were arranged inside the nanotube in a columnar stack with a distance of 4 {\AA} between
two neighboring molecules as shown in Figure 1(a). We prepared a series of systems as described above with CNT diameter ranging from 1.40 nm to 1.72 nm.
After initial energy minimization, systems were gradually heated from 0 K to 300 K. A production run of 50 ns was
then performed in the NVT ensemble. During the simulation, the CNT was held fixed with its axis along the $z$ direction. All molecular dynamics simulations were performed using AMBER package\cite{pearl} with GAFF\cite{wang}
force field. 
Equilibrated simulation snapshot of the systems are shown in Figure 1(b). For the nanotube diameter between 1.40 (n=18, m=0) and 1.56 nm
(n=20, m=0) the coronene molecules maintain well order columnar structure inside CNT. For the diameter larger than that no well defined column formed inside the CNT (see Figure 1(b) and Figure S6 of the supplementary information). Please note in bulk, the coronene molecules do not form any  thermodynamically stable 
ordered structure\cite{cher}. 

\begin{figure}
%%\begin{subfigure}
\includegraphics[scale=0.28]{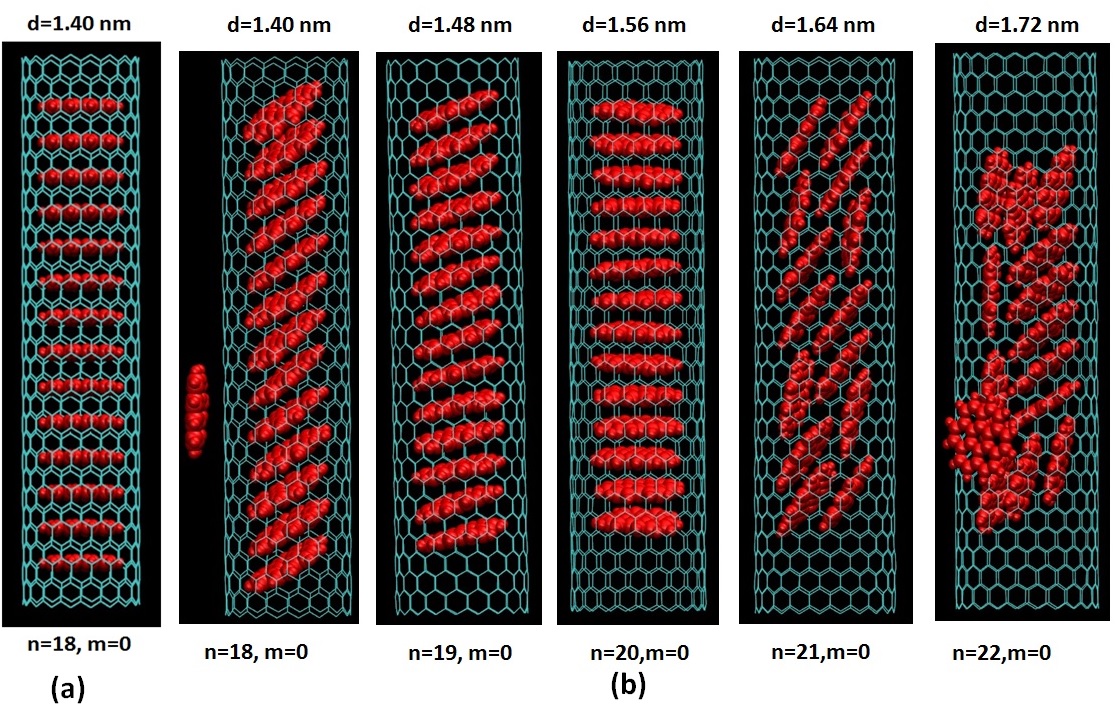}
%\caption{A subfigure}
%\end{subfigure}
%\begin{subfigure}
%\includegraphics[scale=0.3]{all.jpg}
%\caption{A subfigure}
%\end{subfigure}
\caption{(a)Snapshot of the initial arrangement of the coronene molecules inside the CNT (n=18, m=0). 
(b) Equilibrated simulation snapshot of the CNT encapsulated coronene system for the CNTs of diameter ranging from 1.4 nm-1.72 nm. For the CNTs with diameter between 1.40 and 1.56 nm  nice coronene column exists inside the nanotube. For larger diameter, no well order column formed inside CNT.} 
\end{figure} 

Next we measure the positional and orientational correlation between the molecules stacked inside the CNT, since it determines the
 efficiency of charge transport through the coronene stack by controlling the $\pi-\pi$ electronic overlap between the molecules. The
  pair correlation function of the coronene molecules stacked inside the nanotube was caculated using, 
\begin{equation}
  g(z)=\frac{1}{N}\sum_{i=1}^{N}\sum_{j=1,j\neq i}^{N} \left\langle \delta(z-z_{ij})\right\rangle 
\end{equation}
where $z_{ij}$ is the axial separation between the $i$th and the $j$th coronene molecules, $N$ is the number of coronene molecules inside
the nanotube, and the angular brackets indicate an average over time. The resultant plot for (20,0) CNT case is shown in Figure 2(a). The 
pair correlation function for other cases are shown in Figure S1 of the supplementary information.   
The sharp peak in the correlation function indicate almost defect free arrangements of the coronene molecules inside the nanotube. 
Average distance between the two neighbouring coronene molecules is 3.6 {\AA}  for (20,0) CNT with a diameter of 1.56 nm. Since the
$\pi - \pi$ overlap between two coronene molecules decreases exponentially with the distance\cite{feng}, the charge transport 
through the coronene column encapsulated between the (20,0) CNT are expected to be most efficient.

\begin{figure}
\centering
\includegraphics[scale=0.3]{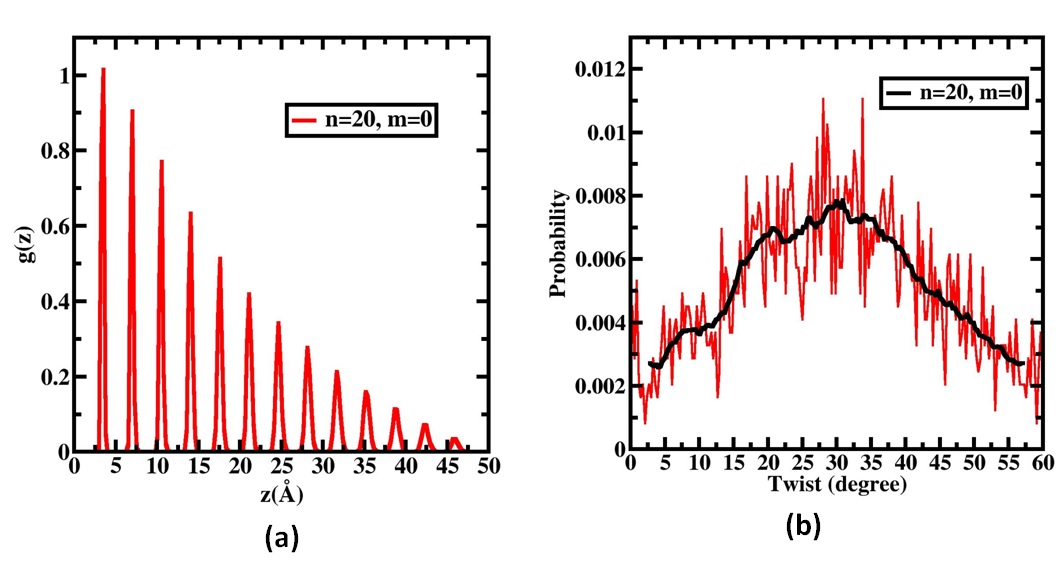}
%%\subfloat[]{\includegraphics[scale=0.12]{cor_19_0.jpg}}
%%\subfloat[]{\includegraphics[scale=0.12]{cor_20_0.jpg}}
\caption{The pair correlation function (a) of  the coronene molecules and twist (b) between the neighboring 
 coronene molecules encapsulated inside (20,0) CNT.  The sharp peak in the
 correlation function (a) indicate almost defect free arrangements of the coronene molecules inside CNT. The molecules are preferred to 
 have a twist angle of $30^{\circ} $ between them(b). The original data is shown in  
 red while a running averaged data in black is shown to  guide the eyes (b).} 
%\label{fig:EcUND} 
\end{figure} 

At this point we ask the question whether the correlation between the coronene molecules as described above depends on the length of the
column inside the nanotube. To answer this question we simulate a long coronene stack of 42 molecules inside the (20,0) CNT. We measure the pair correlation using the same equation 1, but taking only the 14 molecules in the central region of the CNT to
compare this with the correlation of the coronene molecules in the smaller stack (14 molecules).   
We found (see Figure S2 of the supplementary information) identical correlation  among the coronene molecules for both the cases.
We further fit the correlation function to an exponential distribution 
 (see Figure S3 of the supplementary information) and find a correlation length of ~18 {\AA} between the coronene molecules. We 
also calculate the twist between the neighboring molecules in the column (Figure 2(b)), since the overlap between the $\pi-\pi$
 electronic cloud of the neighboring molecules strongly depend on the twist\cite{feng}. Average twist was found out to 
be $30^{\circ}$ between the molecules for the system with CNT diameter 1.56 nm.
To understand why the molecules
prefer to have a  $30^{\circ}$ twist,  we calculated the energy (using M06-2X/6-311g level of theory) of a pair of coronene as a 
function of the twist angle between them (see figure 4(a)). The energy curve has a minima at an angle of $30^{\circ}$ which explains the tendency 
of the molecule to achieve $30^{\circ}$ twist. However there are no specific twist found (see Figure S4 of the supplementary
information) between the coronene molecules confined in (18,0) and (19,0) CNT. In these cases due to smaller radius of the 
CNT the coronene molecules are tilted to accomodate themselves inside CNT and do not have 
the rotaional degree of freedom to reach to the state with $30^{\circ}$ twist.

%%\begin{figure*}
%%\centering
%%\subfloat[]{\includegraphics[scale=0.12]{twist_18.jpg}} 
%%\subfloat[]{\includegraphics[scale=0.12]{twist_19.jpg}}
%%\subfloat[]{\includegraphics[scale=0.12]{twist_20.jpg}}
%%\caption{Twist between the neighboring coronene molecules encapsulated inside CNT of various diameters. The original data is shown in 
%%red while a running averaged data in black is shown to guide the eyes.} 
%\label{fig:EcUND} 
%%\end{figure*} 

We also measured the tilt of the molecules with respect to the axis of the CNT (see Figure S5 of the supplementary information)
which completes the description of the arrangement of coronene molecules inside CNT. Table 1 tabulates numerical values of  all the
quantities describing the arrangement of the coronene molecules inside the CNTs of various diameter and chirality. 

Now we ask the question how can one achieve such system in laboratory. 
%Now we want to convince the reader about the fact that the system under consideration is not a hypothetical and idealized one but can
%be realized very easily in experiment. 
We perform another simulation with the CNT  placed in a box of coronene molecules either in liquid or gas phase as shown in Figure 3(a). The coronene molecules spontaneously entered inside the CNT and form well ordered column
(Figure 3(b)) within few picoseconds simulation run. In this case also we followed the same simulation methodology as described in
the previous paragraph. The two different systems with coronene  molecules kept outside CNT and the coronene molecule arranged 
inside CNT lead to same arrangements of the molecules after equilibration. This simulation suggests an experimental setup  in which 
the CNT is placed inside the coronene vapor to achieve the nanotube encapsulated coronene stack.
\begin{figure}
\centering
\includegraphics[scale=0.25]{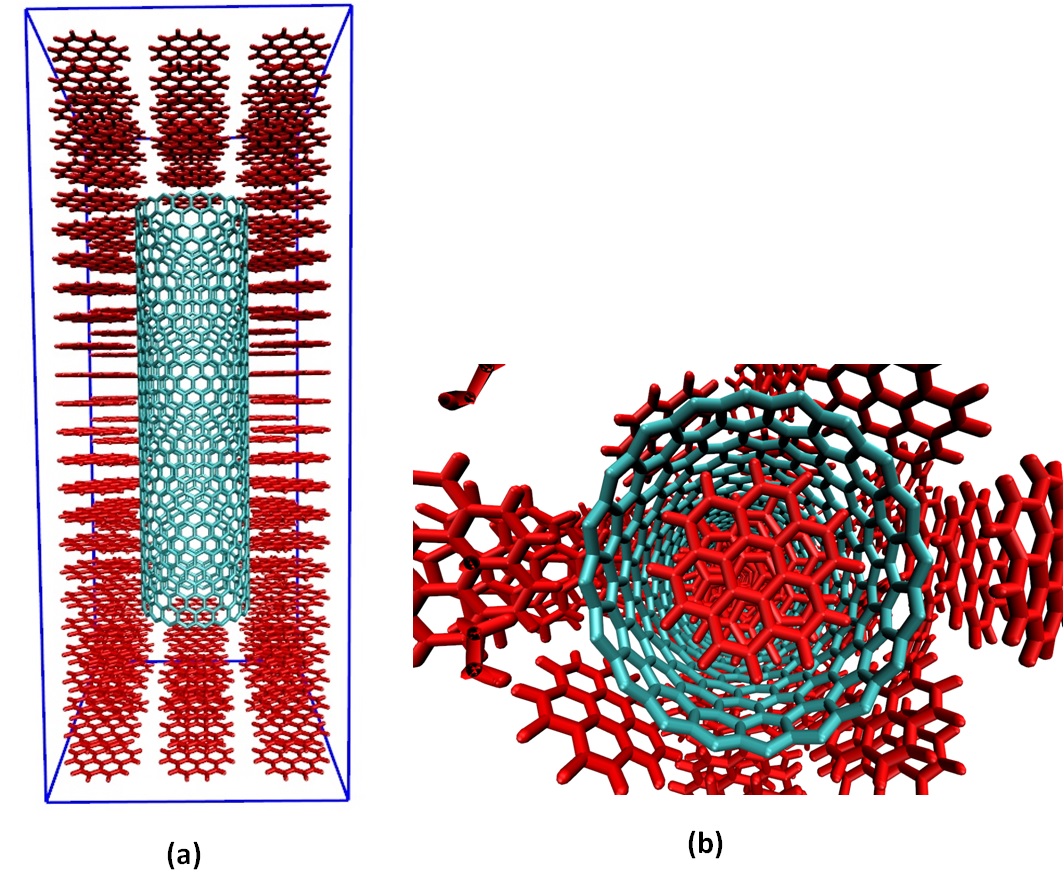}
%%\subfloat[]{\includegraphics[scale=0.15]{final.jpeg}}
\caption{Initial arrangements of the system with CNT placed in a box of coronene molecules (a). Equilibrated snapshot of the system (b).
 Coronene molecules are spontaneously inserted into the nanotube and form column} 
%%\label{fig:EcUND} 
\end{figure} 

\begin{center} {\bf Charge Transport Simulation } \end{center}

We now calculate the charge carrier mobility through the columnar stack of coronene encapsulated inside CNT as obtained from the MD
simulations. To calculate the mobility we use semi-classical Marcus-Hush formalism\cite{mar} which has been extensively used in the
past in various types of organic electronic materials including the discotics. 
In the framework of Marcus-Hush formalism charge transport is described as the thermally activated hopping mechanism between the hopping
sites $i$ and $j$ with the rate
\begin{equation}
 \omega _{ij}=\frac{|{J_{ij}}|^{2}}{\hbar}\sqrt{\frac{\pi}{\lambda k T}}\exp{[-\frac{(\Delta G _{ij}-\lambda )^{2}}{4\lambda kT}]}
\end{equation}
Here $J_{ij}$ is the transfer integral defined as $J_{ij}=\left\langle \Phi _{i}|\hat{H}|\Phi _{j}\right\rangle$. 
$\Phi _{i}$ and $\Phi _{j}$ are the diabatic wave function localized on the $i$th and $j$th hopping site 
respectively. $\Delta G_{ij}$ is the free energy difference between the hopping sites. $\lambda $ is the reorganization
energy of charge transfer, $k$ is the boltzman constant and $T$ is the temperature.\
In our case charge hopping sites are the coronene molecules itself. So we shall use the term hopping site and the coronene molecule  
interchangeably. Highest occupied molecular orbital (HOMO) and lowest unoccupied molecular 
orbital (LUMO) were taken as diabatic wave function to calculate  the transfer integral $J_{ij}$ for the hole and electron transport 
respectively.
Since the HOMO and LUMO orbital of the coronene molecule is degenerate due to 
molecular symmetry of the coronene we calculte four terms $\left\langle HOMO ^{i}|\hat{H}|HOMO ^{j}\right\rangle$, 
$\left\langle (HOMO-1) ^{i}|\hat{H}|HOMO ^{j}\right\rangle$, 
$\left\langle HOMO ^{i}|\hat{H}|(HOMO-1) ^{j}\right\rangle$ and $\left\langle (HOMO-1) ^{i}|\hat{H}|(HOMO-1) ^{j}\right\rangle$. 
$J _{ij}$ (for hole transport) was taken as the root mean square of these four terms. To calculate these terms we performed density 
functional theory calculations (at the level of B3LYP/6-311g) using Gaussian 09\cite{gau} along with the code avaiable in 
VOTCA-CTP module\cite{ruh}. For an application of electric field $F$ 
along the CNT axis the 
free energy difference $\Delta G_{ij}$ turned out to be $F\cdot d _{ij}$. 
Here $d_{ij}$ is the displacement vector between the centre of mass of two coronene molecules. The reorganization $\lambda $ was
calculated using 
\begin{equation}
\lambda = U^{nC}-U^{nN}+U^{cN}-U^{cC}
\end{equation}
Here $U^{nC} (U^{cN})$ is the internal energy of the neutral (charged) coronene molecule with charged (neutral) state geometry 
and $U^{nN}(U^{cC})$ is the internal energy of the neutral (charged) molecule with neutral (charged) state geometry. 
DFT calculations using B3LYP-6-311G level of theory were performed to calculate all these energy terms in equation 3.  

Once all the rates $\omega$ between all the neighbouring coronene molecules was calculated, then Kinetic Monte Carlo (KMC) simulation
was done to simulate transport of charge and the charge mobility was predicted. Charge transport simulation was repeated with 30 
different morphology of the coronene column as obtained from the MD simulation.    

More details of the mobility calculation using the Marcus-Hush formalism can be found in the reference \cite{bag}, \cite{kirk} 
and \cite{ruh}. We find a hole mobility of 14.93 cm$^2$V$^{-1}$s$^{-1}$ at an electric field of 10$^{7}$ V/m 
(applied along the CNT axis) and at a temperature of 300 K for the coronene column encapsulated inside CNT of diameter 
1.56 nm (n=20, m=0). For the systems with CNT diameters 1.40 and 1.48 nm  , the mobility values were found out to be 4.21 and 
4.52 cm$^2$V$^{-1}$s$^{-1}$  respectively.  The mobility (14.93 cm$^2$ V$^{-1}$s$^{-1}$) that we report here is the two order of
magnitude higher than the reported mobility values in discotics\cite{feng}. The reason behind such dramatically high 
mobility lies in the form of pair correlation function itself. The sharply peaked correlation function indicates solid 
like ordering of the molecule inside the nanotube which gives rise to a very sharply peaked distribution of transfer 
integral (Figure 4(b)) resulting such high value of mobility\cite{bag}. As the diameter of the CNT is decreased from 1.56 nm,  the average value of 
the transfer integral also decreases resulting  reduction in the mobility value .  Average value of the transfer integral and 
the mobility values are tabulated in Table 1 for CNTs of various diameter and chirality.

It is worth mentioning that the average  twist between the coronene molecules ($30^{\circ}$)  for the system with highest mobility is not the best
one can get in terms of achieving the highest transfer integral values between the neighboring molecules. A similar arrangement of 
the coronene molecules but with a $60^{\circ}$ twist will lead to much higher mobility. So far we have not found a design principle to 
achieve such twist for coronene.

So far we have talked about the encapsulation of coronenes in zigzag CNT. However there is nothing special about zigzag CNT, it is the radius  of the CNT what is important. 
A (5,17) chiral CNT with the radius 1.56 nm will also gives rise to similar arrangement 
(see Figure S6 of the supplemetary information) of coronene inside CNT and
similar charge carrier mobility as it was for (20,0) CNT case. The possible armchair CNT with radius close to the 1.56 nm are the (11,11)
with diameter 1.49 nm and (12,12) CNT with radius 1.63 nm. None of them allow the best arrangement (see Figure S6 of the supplementary 
information) of the coronene inside themselves to 
achieve the ultrahigh mobility as found in the coronene column inside CNT of diameter 1.56 nm.

\begin{figure}
\includegraphics[scale=0.25]{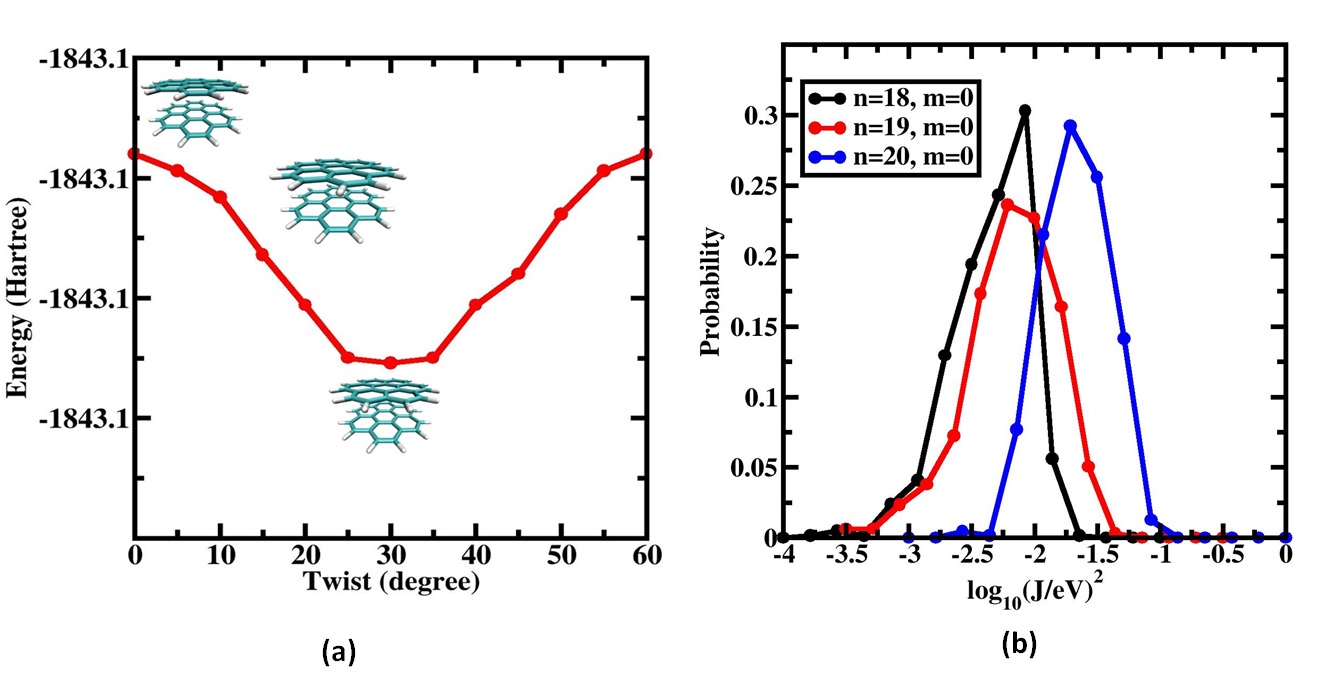}
\caption{Energy of a pair of coronene molecules as a function of the twist angle between them (a). Snapshots of the coronene pair at different twist are shown in the inset of the figure (a). Probability distribution of the transfer integral between the neighboring coronene molecules encapsulated inside CNT.}
\end{figure}

\begin{table*}
\caption{Structural and charge transport properties of the coronene column formed inside CNT of different diameters. }
\begin{ruledtabular}
\begin{tabular}{ | p{2cm} | p{1.4cm} | p{3cm} | p{3cm} | p{2.45cm} | p{2.3cm} | p{2cm}|}
CNT Chirality & CNT Diameter (nm) &Average Distance  between the coronene molecules ({\AA}) &Average twist between the coronene 
molecules (degree)
  &Average Tilt of the coronene molecules (degree) &Average value of the transfer integral (eV) & Hole Mobility (cm$^2$V$^{-1}$s$^{-1}$)\\
\hline
n=18, m=0  & 1.40 & 4.30 & None & 35 & 0.080 &4.21 $\pm$ 0.17\\
n=19, m=0 & 1.48 & 3.80 & None & 20 & 0.099 &4.52 $\pm$ 0.34\\
n=20, m=0 & 1.56 & 3.60 & 30 & 4 & 0.172 & 14.93 $\pm$  0.49\\
n=11, m=11 & 1.49 & 3.65 & None & 18 & 0.109 & 6.49 $\pm$  0.49\\
n=5, m=17 & 1.56 & 3.60 & 30 & 4 & 0.159 & 14.41 $\pm$  0.57\\

\end{tabular}
\end{ruledtabular}
\end{table*}

\begin{center} {\bf Conclusions } \end{center}

The coronene molecules spontaneously self assemble inside CNTs of different chirality. For a specific diameter of 1.56 nm which 
corresponds to either (20,0) zigzag or (5,17) chiral tube we achieve solid like ordering with
average spacing of 3.6 {\AA}  between the coronene molecules. In the confined state we see the  crystal-like local positional ordering of the
molecules which gives rise to a highly efficient charge transport pathway through the coronene column. We report a record charge carrier
mobility of 14.93 cm$^2$V$^{-1}$s$^{-1}$ in this system which is two order of magnitude higher than the reported mobility in the
discotics so far\cite{feng}. We have given proof of principle demonstration to achieve such ordering inside CNT. 
The reported system should  be easily achievable in the experiment and successful use of it will 
lead it to a new frontier in energy application.  

\begin{center} {\bf Acknowledgements } \end{center}
We acknowledge financial support from DST, India. We acknowledge Prof. K. Ganapathy Ayappa for careful reading of the manuscript. We thank Prof. Manish Jain for helpful discussions.

\end{document}

% The following information is for internal review, please remove them for submission
%\widetext
%\leftline{Version 04 as of \today}

% the following line is for submission, including submission to the arXiv!!
%\hspace{5.2in} \mbox{Fermilab-Pub-04/xxx-E}

\title{Ultrahigh Charge Carrier Mobility in Nanotube Encapsulated  Coronene Stack : Supplementary information}
\author{Saientan Bag}
%\email{xxx}
\affiliation{Center for Condensed Matter Theory, Department of Physics, Indian Institute of Science, Bangalore-560012, India}
\author{Prabal K. Maiti}
\email{maiti@physics.iisc.ernet.in}
\affiliation{Center for Condensed Matter Theory, Department of Physics, Indian Institute of Science, Bangalore-560012, India}      
\date{\today}

\maketitle

\begin{figure}[htp]
\includegraphics[scale=.4]{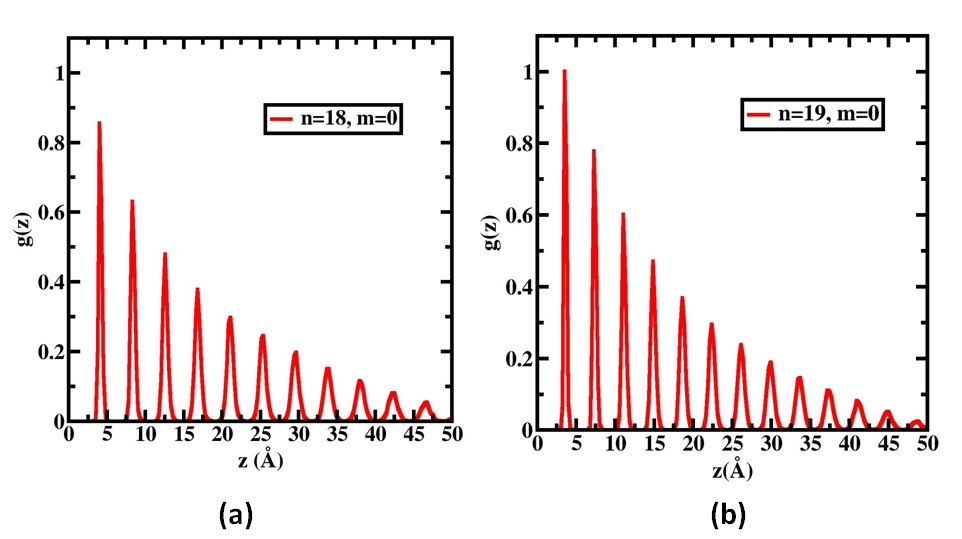}

%\captionsetup[figure]{labelformat=empty}

\caption{Pair correlation function of the coronene molecules encapsulated inside the (18,0) CNT (a) and (19,0) CNT (b). } 
\end{figure} 

\begin{figure}[htp]
\includegraphics[scale=.5]{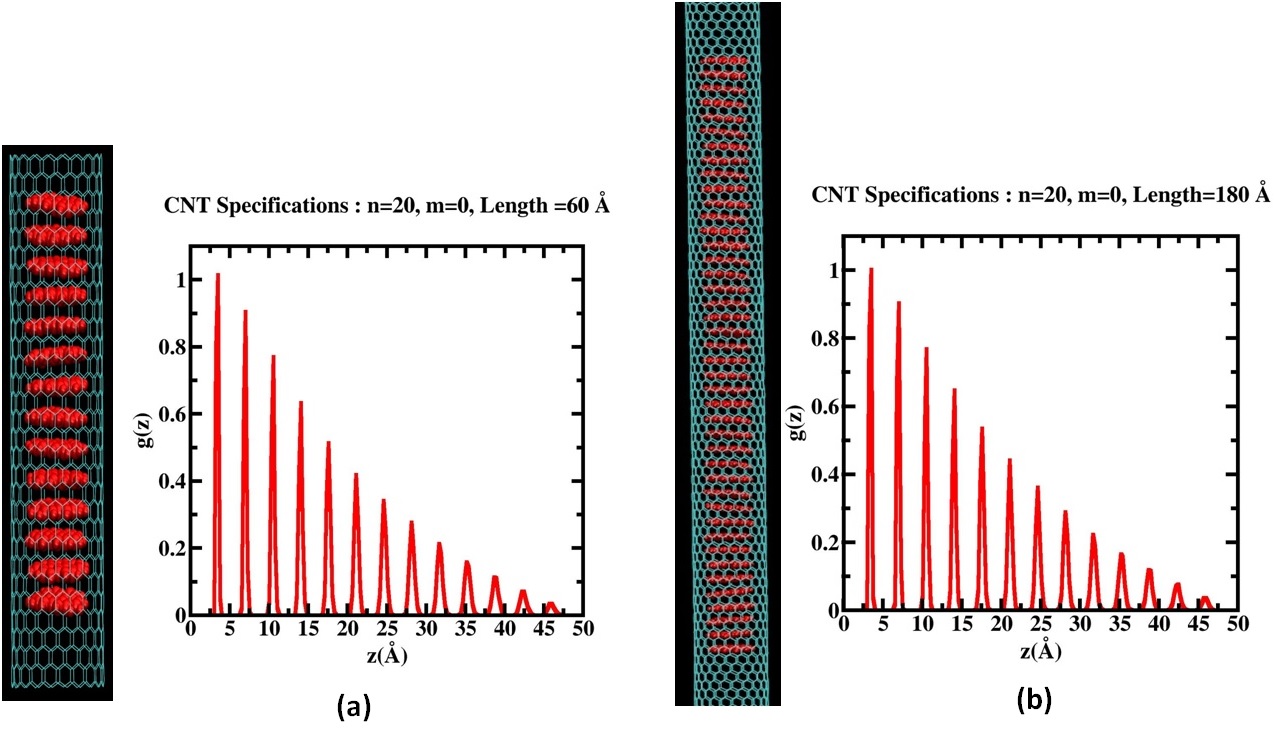}
\caption{Pair correlation function of the coronene molecules encapsulated inside CNT. Equilibrated snapshots of the system are shown in the left of each figure.  Pair correlation function remains unaltered as the length of the column increases.}
%\caption{ Pair correlation function (in red) of the coronene molecules encapsulated inside cnt of radius 1.56 nm (n=20, m=0) and exponential fit of the pair correlation function (in blue). } 
 \end{figure} 

\begin{figure}[htp]
\includegraphics[scale=.14]{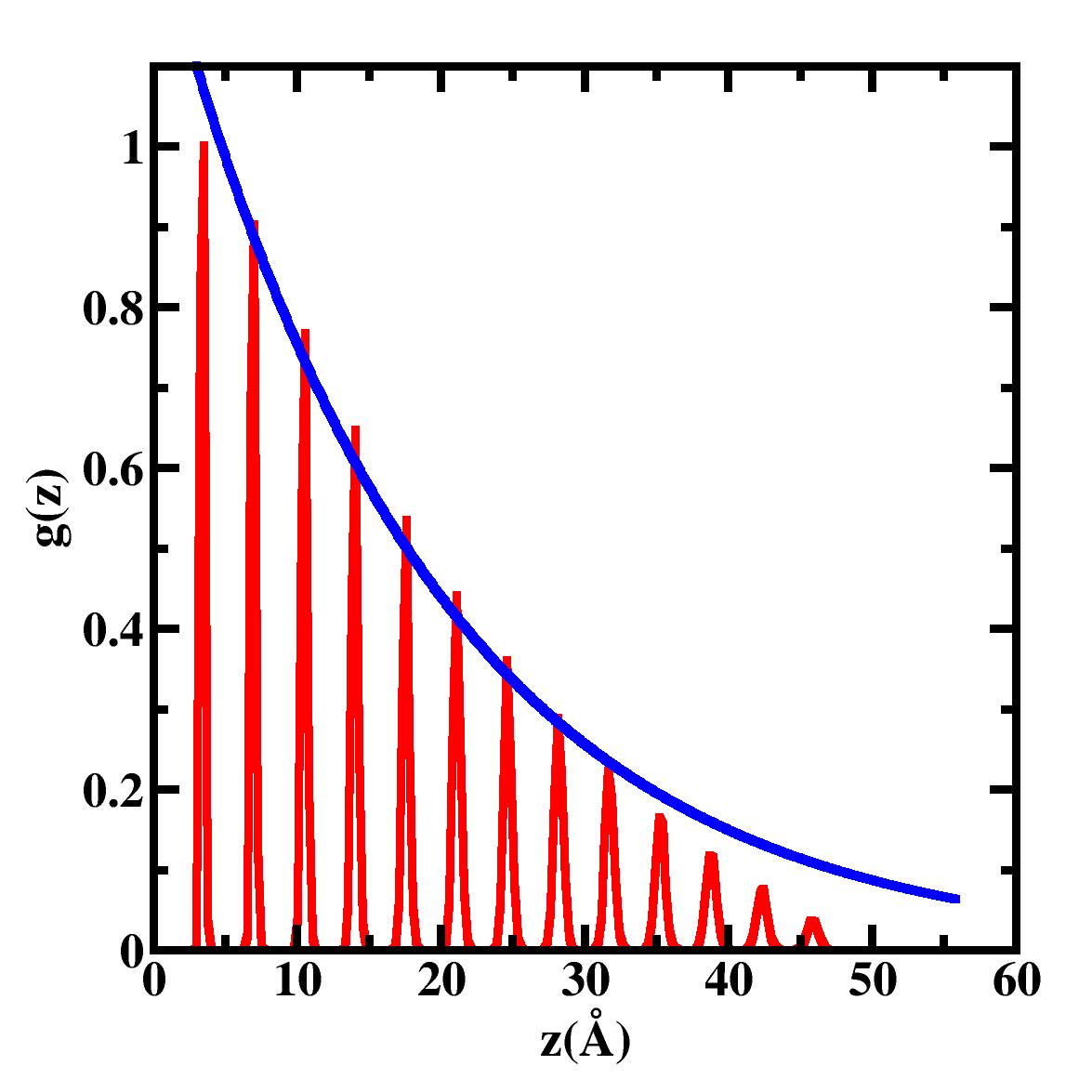} 

\caption{ Pair correlation function (in red) of the coronene molecules encapsulated inside CNT of radius 1.56 nm (n=20, m=0) and exponential fit of the pair correlation function (in blue). } 
 \end{figure} 

\begin{figure}[htp]
\includegraphics[scale=.5]{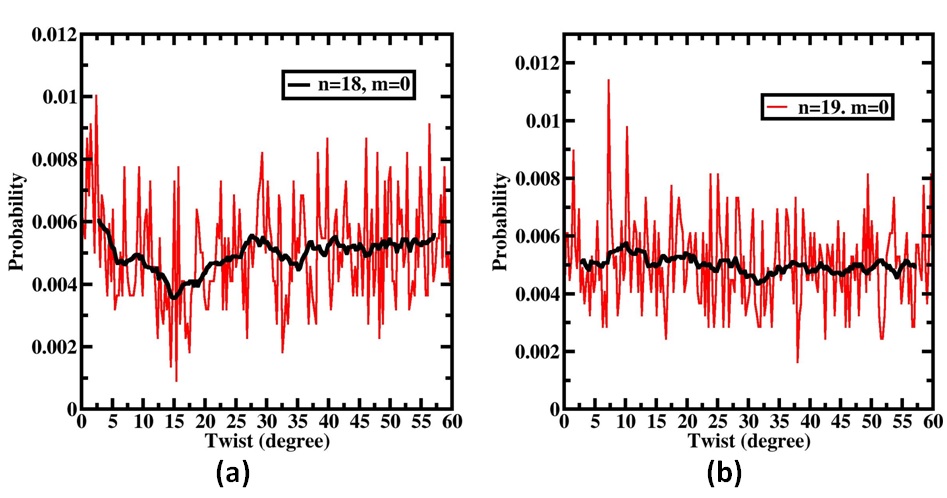}
\caption{Twist between the neighboring coronene molecules encapsulated inside (18,0) and (19,0) CNT. The original data is shown in 
red while a running averaged data in black is shown to guide the eyes.}
\end{figure}

\begin{figure}[htp]
\includegraphics[scale=.15]{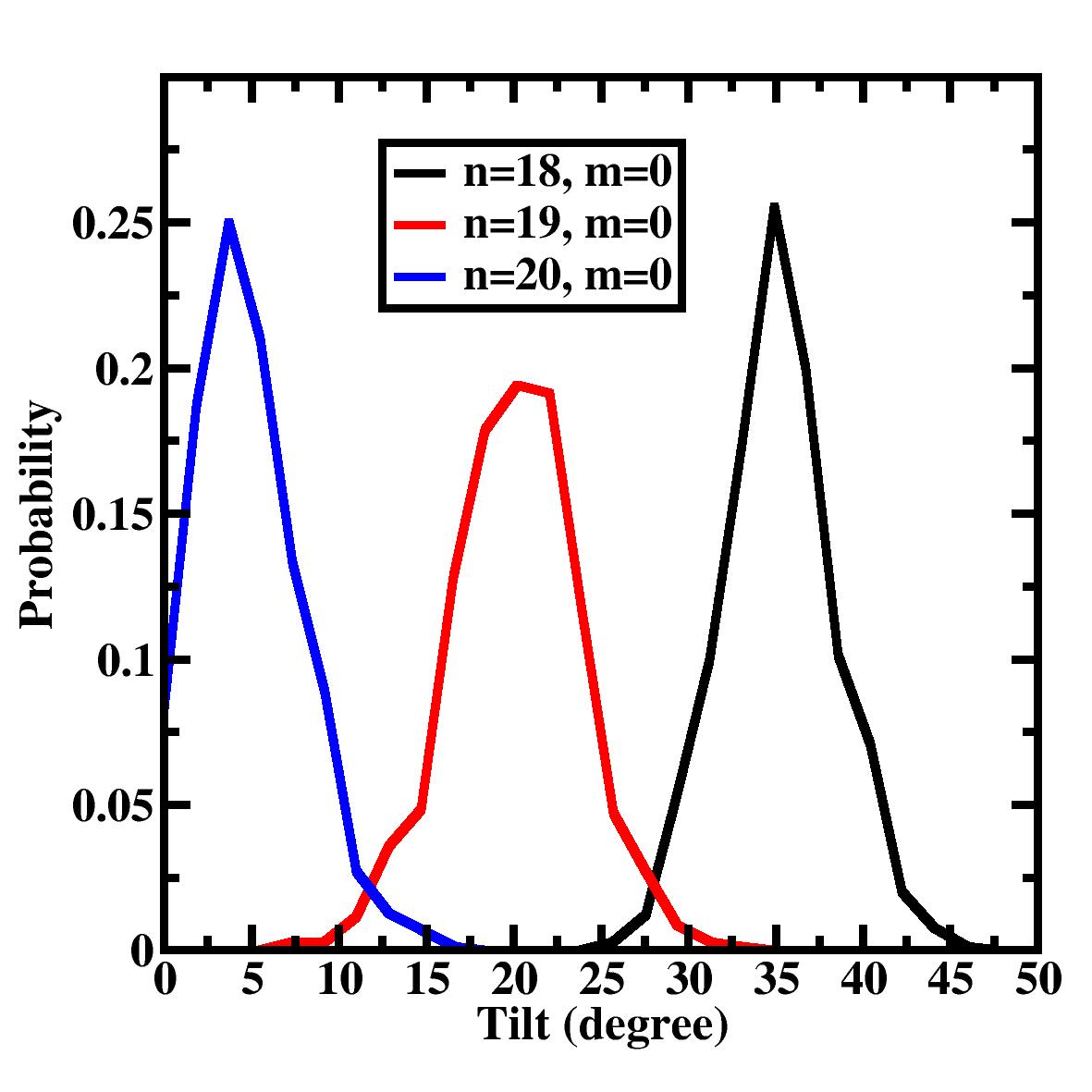}

\caption{ Probability distribution function of the tilt (angle between the CNT axis and the normal of the coronene molecule) of coronene molecules encapsulated inside CNT. Tilt angle is maximum for the case of system with CNT diameter 1.40 nm (n=18, m=0) and it gradually decreases as the CNT diameter increases. } 
 \end{figure}

\begin{figure}[htp]
\includegraphics[scale=.5]{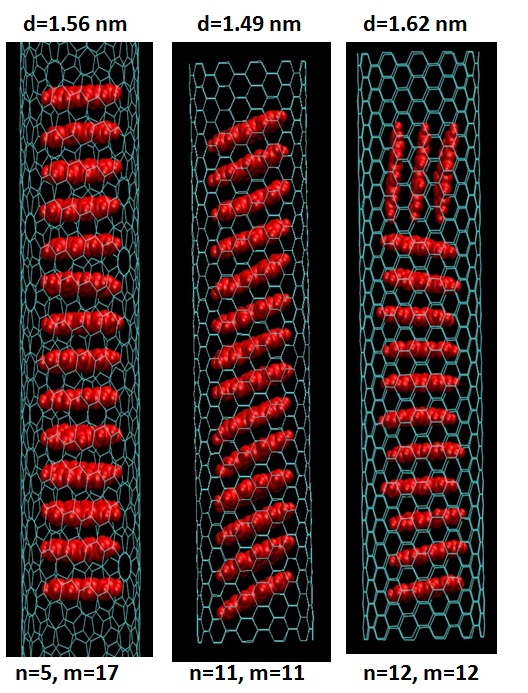}
\caption{Equilibrated simulation snapshot of the various CNT encapsulated coronene systems.}
\end{figure}

%%%%%%%%%%%%%%%%%%%%%%%%%%%%%%%%%%

%%%%%%%%%%%%%%%%%%%%%%%%%%%%%%%%%%%%%%%

%%%%%%%%%%%%%%%%%%%%%%%%%%%%%%%%%%%%%%